# Magnetophoretic separation of blood cells at the microscale

E. P. Furlani

*Institute for Lasers, Photonics and Biophotonics,*
*University at Buffalo (SUNY), Buffalo, NY, 14260*

## Abstract

We present a method and model for the direct and continuous separation of red and white blood cells in plasma. The method is implemented at the microscale using a microfluidic system that consists of an array of integrated soft-magnetic elements embedded beneath a microfluidic channel. The microsystem is passive, and is activated via application of a bias field that magnetizes the elements. Once magnetized, the elements produce a nonuniform magnetic field distribution in the microchannel, which gives rise to a force on blood cells as they pass through the microsystem. In whole blood, white blood cells behave as diamagnetic microparticles while red blood cells exhibit diamagnetic or paramagnetic behavior depending on the oxygenation of their hemoglobin. We develop a mathematical model for predicting the motion of blood cells in the microsystem that takes into account the dominant magnetic, fluidic and buoyant forces on the cells. We use the model to study red/white blood cell transport, and our analysis indicates that the microsystem is capable of rapid and efficient red/white blood cell separation.

# I. INTRODUCTION

Magnetophoresis involves the manipulation of magnetic particles in a viscous medium using an applied magnetic field [1]. Research in this area has intensified recently, with an emphasis on applications in microbiology and biomedicine. Much of this work has focused on the development of microfluidic systems with magnetic functionality that can process magnetically tagged biomaterials such as cells, enzymes, antigens, and DNA [2,3]. Magnetophoretic microsystems are well suited for bioapplications because they enable (i) fast reaction times, (ii) the analysis and monitoring of small samples (picoliters), and (iii) the integration of "micro total analysis systems" (μTAS). However, while numerous microsystems have been developed for processing magnetically tagged biomaterial, relatively little work has been done on the direct magnetophoretic separation of blood cells at the microscale. Indeed, few such systems have been reported, despite the substantial need for fast, accurate, and inexpensive blood cell analysis [4,5].

In whole blood, white blood cells (WBC) behave as diamagnetic microparticles, while red blood cells (RBC) exhibit diamagnetic or paramagnetic behavior depending on whether they are oxygenated or deoxygenated, respectively [6]. Thus, the magnetic force on white blood cells is opposite to that on deoxygenated red blood cells.

In this paper, we present a novel method and a mathematical model for the direct and continuous separation of red and white blood cells in plasma. The method involves the use of a passive magnetophoretic microsystem that consists of an array of integrated soft-magnetic elements embedded beneath a microfluidic channel. The magnetic elements, which are magnetized by a bias field, produce a nonuniform field distribution that gives rise to a magnetic force on blood cells as they flow through the microchannel (Fig. 1). The microsystem is oriented with the fluid flow parallel to the gravitational force, and the cells are separated perpendicular to the flow. The reason for this orientation is that the fluidic and gravitational forces are stronger than the magnetic force, and need to be orthogonal to it to enable efficient magnetic separation (Fig. 1c).

The mathematical model we develop predicts the motion of blood cells in the microsystem. It takes into account the magnetization of the soft-magnetic elements, and the dominant magnetic, fluidic and buoyant forces on the cells. We use it to study the motion of white blood cells and deoxygenated red blood cells, and our analysis indicates that the magnetic force is sufficient to separate the two types of cells as they flow through the microchannel.

The cell separation method presented here has significant and distinct advantages over competing techniques such as centrifuging, or magnetophoresis that involves magnetically labeld materials. First, blood cells can be continuously separated in their native state, without the need for magnetic tagging. Second, small sample volumes can be processed, with efficient cell separation completed within minutes. Third, a high degree of system integration is possible, which enables the potential for a comprehensive total analysis microsystem (μTAS). Fourth, the microsystem is passive and highly portable. In this regard, note that the cell separation itself consumes no energy, although energy is needed to sustain blood flow through the microsystem. For these reasons, the method and model presented in this paper should stimulate and enable the development of novel microsystems for processing blood cells for a variety of research and diagnostic applications.

## II. Theory

### A. Equations of motion

We develop a model for predicting the motion of blood cells in the microsystem, and then apply the model to study cell separation. We predict cellular motion using Newton's law,

$$m_c \frac{d\mathbf{v}_c}{dt} = \mathbf{F}_m + \mathbf{F}_f + \mathbf{F}_g, \tag{1}$$

where $m_c$ and $\mathbf{v}_c$ are the mass and velocity of the cell, and $\mathbf{F}_m$, $\mathbf{F}_f$, and $\mathbf{F}_g$ are the magnetic, fluidic, and gravitational force (including buoyancy), respectively. The magnetic force is obtained using an "effective" dipole moment approach and is given by [7]

$$\mathbf{F}_m = \mu_0 V_c \left(\chi_c - \chi_f\right)\left(\mathbf{H}_a \bullet \nabla\right)\mathbf{H}_a, \tag{2}$$

where $\chi_c$ and $V_c$ are the susceptibility and volume of the cell, $\chi_f$ is the susceptibility of the transport fluid (in this case plasma), $\mathbf{H}_a$ is the applied magnetic field, and $\mu_0 = 4\pi \times 10^{-7}$ H/m is the permeability of free space. The fluidic force is based on Stokes' law for the drag on a sphere,

$$\mathbf{F}_f = -6\pi \eta R_c (\mathbf{v}_c - \mathbf{v}_f), \tag{3}$$

where $R_c$ is the radius of the cell, and $\eta$ and $\mathbf{v}_f$ are the viscosity and the velocity of the fluid, respectively. The gravitational force is given by

$$\mathbf{F}_g = V_c(\rho_c - \rho_f)g\,\hat{\mathbf{x}}, \qquad (4)$$

where $\rho_c$ and $\rho_f$ are the densities of the cell and fluid, respectively ($g = 9.8$ m/s$^2$). It is important to note that gravity acts in the +x direction, parallel to the flow (see Figs. 1c and 2b).

## B. Magnetic force

The magnetic force on a cell is obtained using an "effective" dipole moment method in which the cell is replaced by an "equivalent" point dipole with a moment $\mathbf{m}_{c,\text{eff}}$ [7]. The force on the dipole (and hence on the cell) is given by

$$\mathbf{F}_m = \mu_f \left(\mathbf{m}_{c,\text{eff}} \bullet \nabla\right) \mathbf{H}_a, \qquad (5)$$

where $\mu_f$ is the permeability of the transport fluid, and $\mathbf{H}_a$ is the applied magnetic field intensity at the center of the cell, were the equivalent point dipole is located. We evaluate Eq. (5) for a cell in a fluid of permeability $\mu_f$ and obtain

$$\mathbf{F}_m = \mu_f V_c \frac{3(\chi_c - \chi_f)}{\left[(\chi_c - \chi_f) + 3(\chi_f + 1)\right]} (\mathbf{H}_a \bullet \nabla) \mathbf{H}_a. \qquad (6)$$

A detailed derivation of Eq. (6) is given in reference [7]. For blood cell separation, $|\chi_c - \chi_f| \ll 1$ and $\mu_f \approx \mu_0$, and therefore Eq. (6) reduces to Eq. (2).

## C. Magnetic field of the magnetized elements

To evaluate the magnetic force, we need an expression for the applied field. This is a superposition of two distinct fields, the bias field $\mathbf{H}_{bias}$, and the field $\mathbf{H}_e$ due to the array of magnetized elements,

$$\begin{aligned}\mathbf{H}_a &= \mathbf{H}_{bias} + \mathbf{H}_e \\ &= \mathrm{H}_{e,x}\hat{\mathbf{x}} + \left(\mathrm{H}_{bias,y} + \mathrm{H}_{e,y}\right)\hat{\mathbf{y}}.\end{aligned} \qquad (7)$$

However, $\mathbf{H}_{bias}$ and $\mathbf{H}_e$ are not both independent. Specifically, $\mathbf{H}_e$ depends on $\mathbf{H}_{bias}$ as it is the bias field that magnetizes the elements. Therefore, $\mathbf{H}_{bias}$ induces $\mathbf{H}_e$. The bias field can be optimized using an analytical formula as described by Furlani [8,9]. Once the bias field is known, we can determine $\mathbf{H}_e$, but for this we need a magnetization model for the magnetic elements.

We use a linear magnetization model with saturation to predict the magnetization of the soft-magnetic elements. Specifically, below saturation,

$$\mathbf{M}_e = \chi_e \mathbf{H}_{in}, \qquad (8)$$

where $\chi_e = \mu_e/\mu_0 - 1$, and $\mu_e$ and the susceptibility and permeability of each element. Above saturation $\mathbf{M}_e = \mathbf{M}_{es}$, where $\mathbf{M}_{es}$ is the saturation magnetization of the element. In Eq. (8) $\mathbf{H}_{in} = \mathbf{H}_a + \mathbf{H}_{demag}$ is the field inside the element. Specifically, $\mathbf{H}_{demag} = -\mathrm{N}_d \mathbf{M}_e$, where $\mathrm{N}_d$ is the demagnetization factor, which is geometry dependent [10]. Thus from Eq. (8) we have $\mathbf{M}_e = \chi_e \left(\mathbf{H}_{bias} - \mathrm{N}_d \mathbf{M}_e\right)$, which can be rewritten as

$$M_e = \frac{\chi_e}{(1+N_d\chi_e)} H_{bias}. \qquad (9)$$

For a soft-magnetic element, $\chi_e \gg 1$ and Eq. (9) reduces to

$$M_e = \frac{H_{bias}}{N_d} \qquad (\chi_e \gg 1). \qquad (10)$$

The demagnetization factor for a highly permeable ($\chi_e \approx \infty$) long rectangular element of width 2w and height 2h that is magnetized parallel to its height can be obtained using analytical formulas (see Fig. 2b). Specifically, both the demagnetization factor $N_d$ and the aspect ratio of the element $p = \frac{h}{w}$ can be defined parametrically as a function of a variable $k$ over the domain $0 < k < 1$ as follows [11]:

$$N_d = \frac{4}{\pi} \frac{\left[E(k) - k'^2 K(k)\right]\left[E(k') - k^2 K(k')\right]}{k'^2}, \qquad (11)$$

$$\frac{h}{w} = \frac{E(k') - k^2 K(k')}{E(k) - k'^2 K(k)}, \qquad (12)$$

where $k' = \sqrt{1-k^2}$, and $K(k)$ and $E(k)$ are the complete elliptic integrals of the first and second kind, respectively,

$$K(k) = \int_0^{\frac{\pi}{2}} \frac{1}{\sqrt{1-k^2\sin^2(\phi)}} d\phi, \qquad E(k) = \int_0^{\frac{\pi}{2}} \sqrt{1-k^2\sin^2(\phi)} \, d\phi. \qquad (13)$$

To determine the magnetization $M_e$ of the elements, we first use Eqs. (11) and (12) to obtain $N_d$ for a give aspect ratio $p$ (see p 191, Table A.2 in reference [11]). Next, we evaluate Eq. (10), taking saturation into taking into account. Specifically, the magnetization of an element is obtained using,

$$M_e = \begin{cases} \dfrac{H_{bias}}{N_d} & H_{bias} < N_d M_{es} \\ M_{es} & H_{bias} \geq N_d M_{es} \end{cases}. \qquad (14)$$

Once $\mathbf{M}_e$ is known, $\mathbf{H}_e$ is easily determined. Specifically, the field solution for a long rectangular element of width 2w and height 2h that is centered with respect to the origin in the x-y plane, and magnetized parallel to its height (along the y-axis as shown in Fig. 2b) is well known (pp 210-211 in reference [10]). The field components are

$$H_{ex}^{(0)}(x,y) = \dfrac{M_e}{4\pi}\left\{\ln\left[\dfrac{(x+w)^2+(y-h)^2}{(x+w)^2+(y+h)^2}\right] - \ln\left[\dfrac{(x-w)^2+(y-h)^2}{(x-w)^2+(y+h)^2}\right]\right\}, \qquad (15)$$

and

$$H_{ey}^{(0)}(x,y) = \dfrac{M_e}{2\pi}\left\{\tan^{-1}\left[\dfrac{2h(x+w)}{(x+w)^2+y^2-h^2}\right] - \tan^{-1}\left[\dfrac{2h(x-w)}{(x-w)^2+y^2-h^2}\right]\right\}. \qquad (16)$$

In these equations, $M_e$ is determined using Eq. (14).

The field and force for an array of elements can be obtained from (15) and (16) [8,9]. Specifically, let $N_e$ denote the number of elements in the array, and let n = (0,1,2,3,4, …, $N_e$ -1) label the individual elements (Fig. 2b). Now, $H_{ex}^{(0)}(x,y)$ and $H_{ey}^{(0)}(x,y)$ denote the field components due to the first element (n=0). The n'th element is centered at $x = s_n$, and its field components can be written as $H_{ey}^{(n)}(x,y) = H_{ey}^{(0)}(x-s_n,y)$ and $H_{ey}^{(n)}(x,y) = H_{ey}^{(0)}(x-s_n, y)$ (see Fig. 2d). The total field of the array is obtained by summing the contributions from all the elements,

$$H_{ex}(x,y) = \sum_{n=0}^{N_e-1} H_{ex}^{(0)}(x-s_n, y), \qquad H_{ey}(x,y) = \sum_{n=0}^{N_e-1} H_{ey}^{(0)}(x-s_n, y). \qquad (17)$$

It follows from Eqs. (2), (7) and (17) that the force components are

$$F_{mx}(x,y) = \mu_0 V_c \left(\chi_c - \chi_f\right) \left[ \left( \sum_{n=0}^{N_e-1} H_{ex}^{(0)}(x-s_n, y) \right) \left( \sum_{n=0}^{N_e-1} \frac{\partial H_{ex}^{(0)}(x-s_n, y)}{\partial x} \right) \right. \\ \left. + \left( H_{bias,y} + \sum_{n=0}^{N_e-1} H_{ey}^{(0)}(x-s_n, y) \right) \left( \sum_{n=0}^{N_e-1} \frac{\partial H_{ex}^{(0)}(x-s_n, y)}{\partial y} \right) \right], \qquad (18)$$

and

$$F_{my}(x,y) = \mu_0 V_c \left(\chi_c - \chi_f\right) \left[ \left( \sum_{n=0}^{N_e-1} H_{ex}^{(0)}(x-s_n, y) \right) \left( \sum_{n=0}^{N_e-1} \frac{\partial H_{ey}^{(0)}(x-s_n, y)}{\partial x} \right) \right. \\ \left. + \left( H_{bias,y} + \sum_{n=0}^{N_e-1} H_{ey}^{(0)}(x-s_n, y) \right) \left( \sum_{n=0}^{N_e-1} \frac{\partial H_{ey}^{(0)}(x-s_n, y)}{\partial y} \right) \right]. \qquad (19)$$

In Eqs. (18) and (19) we have assume that the bias field is constant and in the y-direction. Explicit expressions for the field and force for an array of rectangular soft-magnetic elements (Eqs. (17) - (19)) have been derived and verified using finite element analysis (FEA) [8,9].

### D. Fluidic force

To evaluate the fluidic force in Eq. (3) we need an expression for the fluid velocity $\mathbf{v}_f$ in the microchannel. Let $h_c$ and $w_c$ denote the half-height and half-width of its rectangular cross section (Fig. 2a). We assume fully developed laminar flow parallel to the x-axis and obtain

$$v_f(y) = \frac{3\overline{v}_f}{2}\left[1-\left(\frac{y-(h+h_c+t_b)}{h_c}\right)^2\right], \tag{20}$$

where $\overline{v}_f$ is the average flow velocity and $t_b$ is the thickness of the base of the channel (i.e., the distance from the top of the magnetic elements to the lower edge of the fluid) [8,9]. We substitute Eq. (20) into Eq. (3) and obtain the fluidic force components

$$\mathbf{F}_{fx} = -6\pi\eta R_c\left[v_{c,x} - \frac{3\overline{v}_f}{2}\left[1-\left(\frac{y-(h+h_c+t_b)}{h_c}\right)^2\right]\right], \tag{21}$$

and

$$\mathbf{F}_{fy} = -6\pi\eta R_c v_{c,y}. \tag{22}$$

We use these in the equations of motion below.

### E. Blood cell properties

We need the magnetic properties of white and red blood cells to complete the mathematical model. White blood cells (WBCs) comprise five different kinds of cells that are classified into two groups: agranulocytes (lymphocyte and monocyte), and granulocytes (neutrophil, eosinophil and basophil) [12,13]. The five different cells have different sizes, with diameters that range from 6 μm to 15 μm. We account for the different types of white blood cells by using average WBC properties: $\rho_{wbc} = 1070$ kg/m$^3$, $R_{wbc} = 5$ μm, and $V_{wbc} = 524$ μm$^3$ [12]. White blood cells exhibit a diamagnetic behavior in plasma, but their magnetic susceptibility is not well known [13].

In order to determine the feasibility of WBC separation we use a lower bound estimate for the WBC susceptibility as suggested by Takayasu et al., specifically we use the susceptibility of water $\chi_{wbc} = -9.2 \times 10^{-6}$ (SI) [13]. This value is consistent with measurements made by Han and Frazier in which a value of $\chi_{wbc} \approx -9.9 \times 10^{-6}$ was obtained for WBCs with 5 μm diameters (see Table 1 p 1428 in reference [4]). Thus, the WBC susceptibility that we use provides a conservative lower bound estimate of the force, and enables us to determine the feasibility of WBC separation.

Red blood cells (RBCs), when unperturbed, have a well-defined biconcave discoid shape with a diameter of 8.5 ± 0.4 μm and a thickness of 2.3 ± 0.1 μm. These cells account for approximately 99% of the particulate matter in blood, and the percentage by volume (hematocrit) of packed red blood cells in a given sample of blood, is normally 40-45%. For red blood cells, we use $R_{rbc}$ = 3.84 μm (hydrodynamic radius), $V_{rbc} = 88.4$ μm$^3$, and $\rho_{rbc} = 1100$ kg/m$^3$.[14] The susceptibility of a RBC depends on the oxygenation of its hemoglobin. We use $\chi_{rbc,oxy} = -9.22 \times 10^{-6}$ (SI) and $\chi_{rbc,deoxy} = -3.9 \times 10^{-6}$ (SI) for oxygenated and deoxygenated red blood cells, respectively [13-15]. The transport fluid is plasma, which has the following properties: $\eta = 0.001$ kg/s, $\rho_f = 1000$ kg/m$^3$ and $\chi_f = -7.7 \times 10^{-6}$ (SI) [13-15].

### F. Equations of motion

The equations of motion for blood cell transport through the microsystem can be written in component form by substituting Eqs. (18), (19), (21) and (22) into Eq. (1),

$$m_c \frac{dv_{c,x}}{dt} = F_{mx}(x,y) + V_c(\rho_c - \rho_f)g - 6\pi\eta R_c \left[ v_{c,x} - \frac{3\overline{v}_f}{2}\left[1 - \left(\frac{y-(h+h_c+t_b)}{h_c}\right)^2\right]\right], \quad (23)$$

$$m_c \frac{dv_{c,y}}{dt} = F_{my}(x,y) - 6\pi\eta R_c v_{c,y}, \quad (24)$$

$$v_{c,x}(t) = \frac{dx}{dt}, \qquad v_{c,y}(t) = \frac{dy}{dt}. \quad (25)$$

Equations (23) - (25) constitute a coupled system of first-order ordinary differential equations (ODEs) that are solved subject to initial conditions for $x(0)$, $y(0)$, $v_{c,x}(0)$, and $v_{c,y}(0)$. These equations can be solved numerically using various techniques such as the Runge-Kutta method.

## III. RESULTS

We use the model developed above to study blood cell motion in the microsystem. As a first step, we compute the field due to an array of three magnetized permalloy (78% Ni 22% Fe, $M_{es} = 8.6\times10^5$ A/m [10]) elements (Fig. 3). Each element is 300 mm high, 300 mm wide, and they are spaced 300 mm apart (edge to edge). Thus, w = h = 150 μm, and these elements have an aspect ratio $p = h/w = 1$. From Eqs. (11) and (12) we compute a demagnetization factor of $N_d = 0.456$ (p 191, Table A.2 in reference [11]). The bias field is set to $H_{bias} = 3.9 \times 10^5$ A/m, which from Eq. (14) is sufficient to saturate the elements, i.e. $H_{bias} = N_d M_{es} \rightarrow M_e = M_{es}$. This bias field intensity corresponds to a flux density of 5000 Gauss, which can be obtained by positioning rare earth permanent magnets on either side of the microsystem as shown in Fig. 1.

The field components $B_x$ and $B_y$ due to the magnetized elements are computed along a horizontal line 60 μm above the elements (i.e. at y = 210 μm) using Eqs. (15) - (17) with $N_e = 2$ (Fig. 3). Notice that $B_x$ peaks near the edges of the elements and alternates in sign from one edge to the other, whereas $B_y$ obtains its maximum value at the center of the elements.

Next, we compute the magnetic force on a deoxygenated RBC along the same horizontal line as above (60 μm above the elements). The component $F_{mx}$ acts in the flow direction while $F_{my}$ acts perpendicular to the flow, and is responsible for cell separation. Notice that $F_{mx}$ peaks near the edges of the element, and changes assign across the element (Fig. 4a). Thus, a deoxygenated RBC experiences acceleration in the flow direction as it passes the leading edge of an element, followed by deceleration as it passes the trailing edge. $F_{my}$ is downward (negative) immediately above an element, but alternates in direction across an element, upward to the left of an element, downward above an element, and upward to the right of an element (Fig. 4b). Therefore, a deoxygenated RBC accelerates upward, then downward, and then upward again as it passes an element. Oxygenated RBC and WBC exhibit a similar behavior, but in the opposite direction.

Lastly, we determine the feasibility of RBC/WBC separation by predicting the motion of deoxygenated RBC and WBC as they move through the microsystem. The fluid channel is 120 μm high, 1 mm wide, and 30 mm long, and there are 45 permalloy elements embedded immediately beneath it. Each element is 300 μm high and 300 μm wide, and they are spaced 300 μm apart (edge to edge). Thus, the magnetic element array spans a distance of 26.7 mm along the bottom of the microchannel.

The cells enter the microchannel to the left of the first element ($x(0) = -600$ μm) at various initial heights: y(0) = 165 μm, 180 μm, …, 255 μm. The top of the fluidic chamber is 120 μm above the elements at y = 270 μm. The average fluid velocity is $\bar{v}_f = 0.25$ mm/s, and the cells enter the channel with this velocity. The WBC and RBC trajectories are shown in Fig. 5a and 5b, respectively. The trajectory profiles are irregular due to the spatial variation of the magnetic force as described above. Note that the WBC and RBC separate before they reach the end of the array. Specifically, all WBC move to the top of the channel, while all deoxygenated RBC move to the bottom. The separation times for the WBCs and RBCs are 60s and 80s, respectively (Fig. 6).

The preceding analysis demonstrates the viability WBC/RBC separation. The parameters used in the analysis (e.g. the dimensions and spacing of the magnetic elements) were arrived at through an iterative series of simulations, and do not represent optimum values. However, the model enables rapid parametric analysis, and there are several variables that can be adjusted to optimize performance including the number, size, and spacing of the elements, the dimensions of the microchannel, and the flow rate. Thus, the separation method is robust, and the microsystem holds significant potential for numerous biomedical applications.

## IV. CONCLUSION

We have presented a novel method for the direct and continuous separation of red and white blood cells in plasma that has numerous advantages over existing cell separation methods. The method is implemented in a passive magnetophoretic microsystem that can be fabricated using established methods [4,16,17]. We have also developed a mathematical model for studying blood cell transport at the microscale, and have used the model to predict cell separation in the microsystem. Our analysis indicates that deoxygenated red blood cells can be separated from white blood cells in plasma, and that efficient separation can be achieved within a few minutes. The method and model presented here should stimulate further research into magnetophoretic cell separation, and lead to the development of novel cell separation microsystems.

# Figure Captions

FIG. 1. Magnetophoretic microsystem: (a) microsystem with bias field structure, (b) cross-section of microsystem showing magnetic elements beneath the microchannel, and (c) magnified view of microfluidic channel showing the bias field, magnetic elements, and forces of red and white blood cells (RBC and WBC).

FIG. 2. Magnetophoretic microsystem: (a) microfluidic channel, and (b) cross section of microsystem showing array of magnetized elements.

FIG. 3. Magnetic field above three magnetized elements ($\bullet$ = FEA): (a) $B_x$ (parallel to flow), (b) $B_y$ (perpendicular to flow), (c) three elements embedded beneath flow channel, and (d) cross section of microsystem showing magnetized elements and reference frame.

FIG. 4. Magnetic force on a deoxygenated red blood cell above three magnetized elements ($\bullet$ = FEA): (a) $F_{mx}$ (parallel to flow), (b) $F_{my}$ (perpendicular to flow), (c) three elements embedded beneath flow channel.

FIG. 5. Blood cell trajectories above magnetized elements (upper half of magnetized elements shown for reference): (a) white blood cell (WBC) trajectories, (b) red blood cell (RBC) trajectories.

FIG. 6. Blood cell separation time vs. initial height above magnetized elements: (a) white blood cell (WBC) separations time, and (b) deoxygenated red blood cell (RBC) separation time.

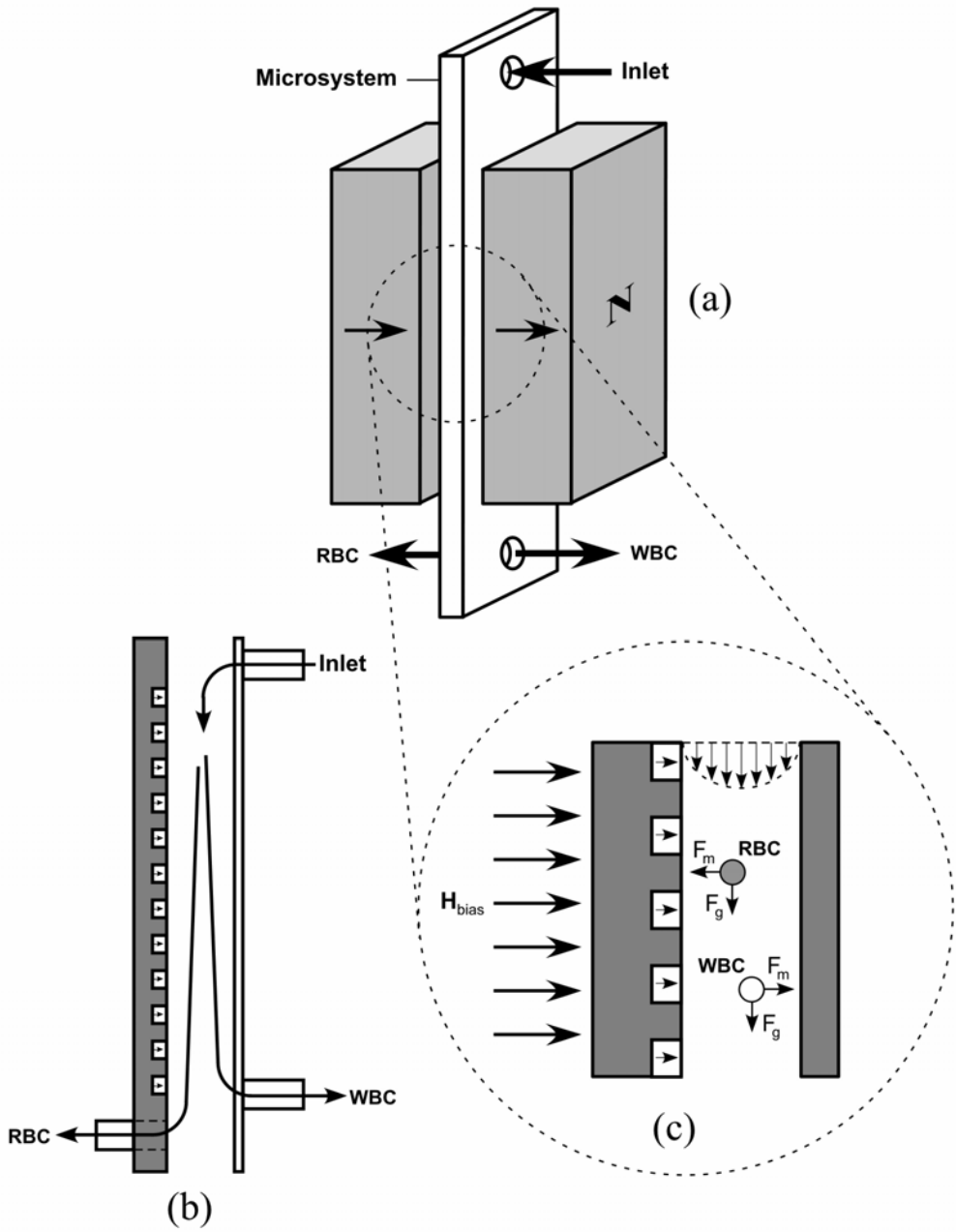

FIG. 1

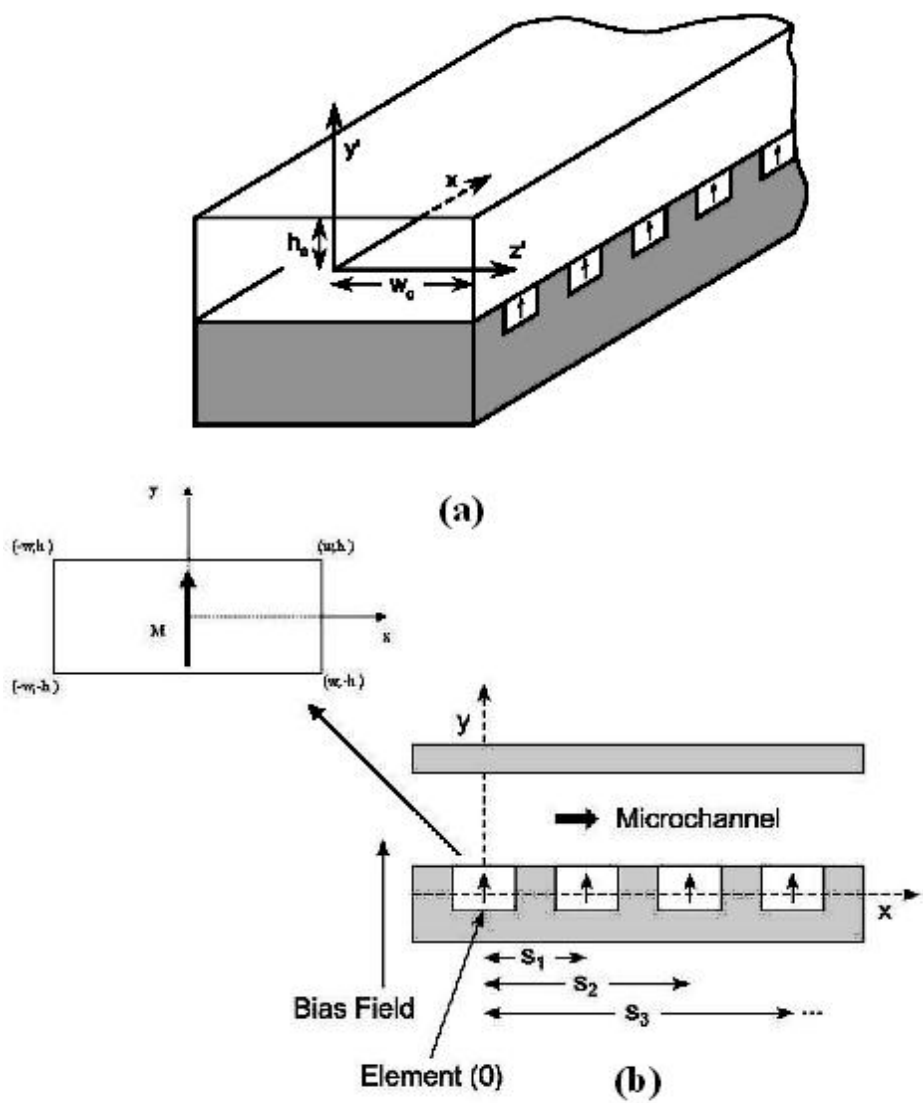

FIG. 2

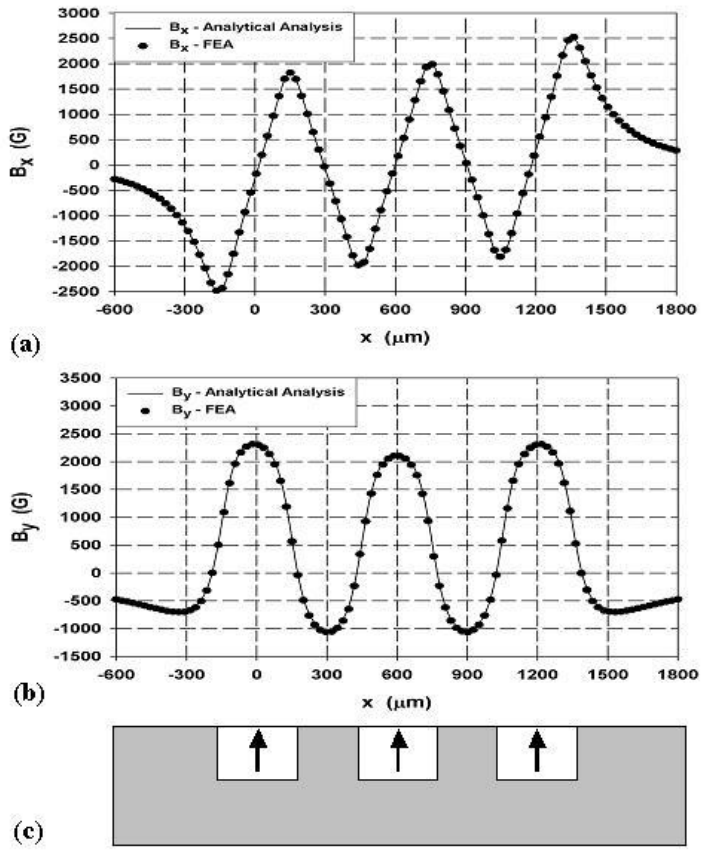

FIG. 3

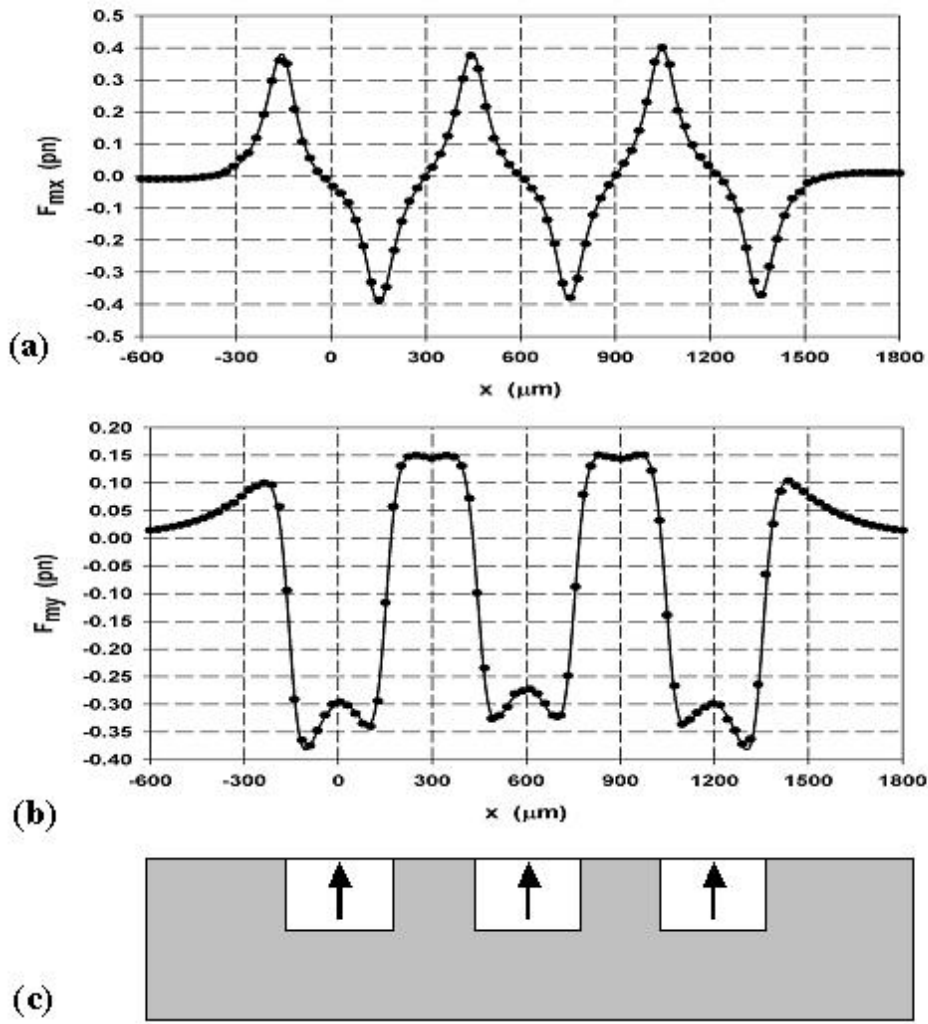

FIG. 4

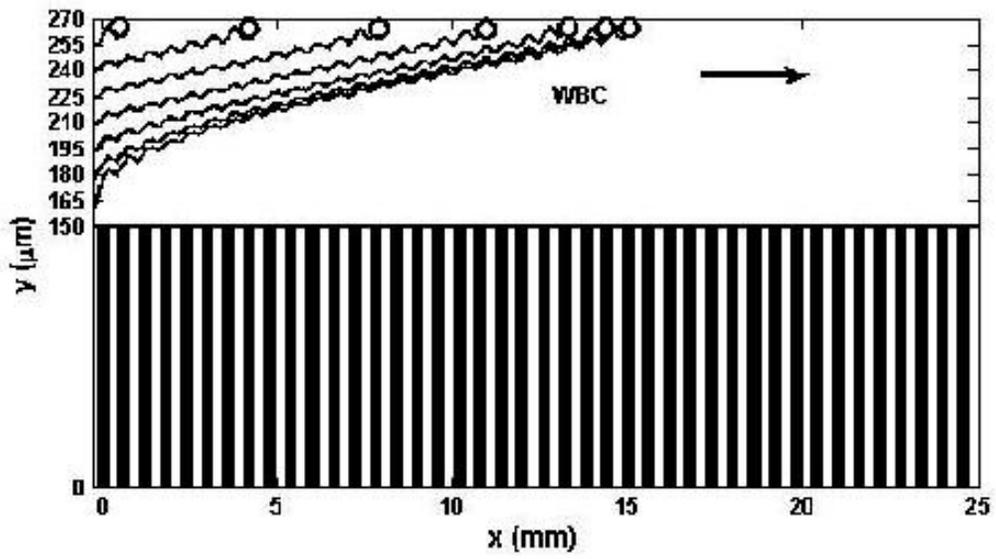

(a)

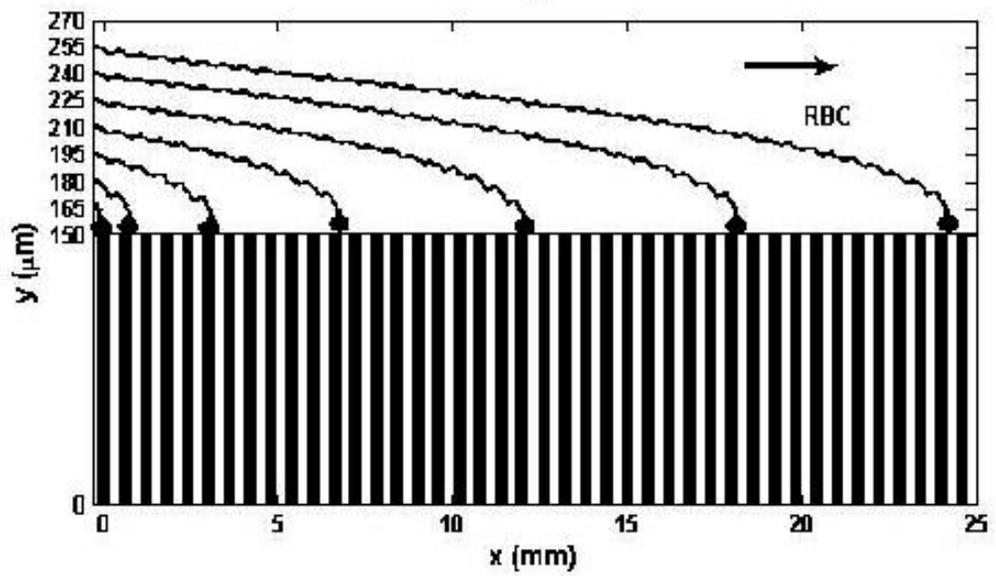

(b)

FIG. 5

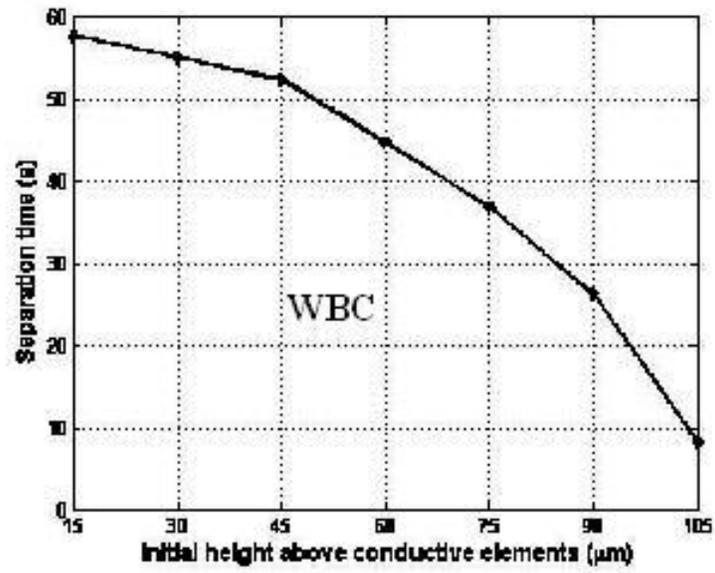

(a)

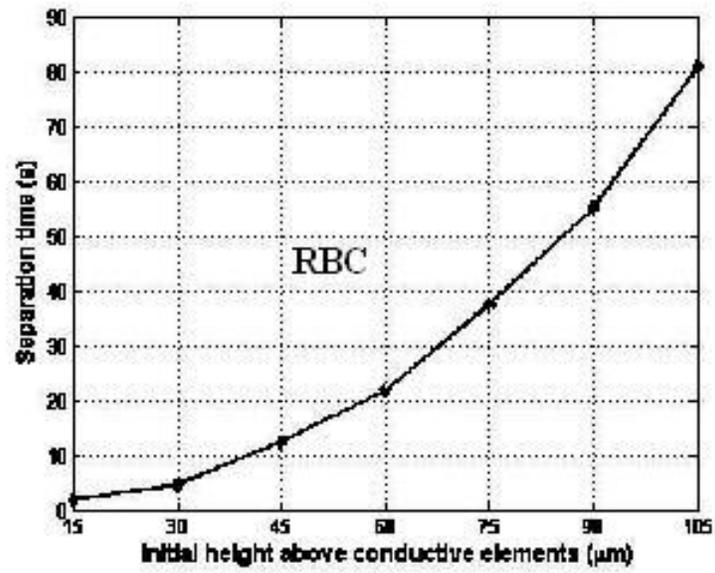

(b)

FIG. 6